\documentstyle[12pt]{article}        
\newlength{\dinwidth}
\newlength{\dinmargin}
\def\lsim{\mathrel{\rlap{\lower4pt\hbox{\hskip1pt$\sim$}}
    \raise1pt\hbox{$<$}}}                
\def\gsim{\mathrel{\rlap{\lower4pt\hbox{\hskip1pt$\sim$}}
    \raise1pt\hbox{$>$}}}                
\newcommand{\be}{\begin{equation}}
\newcommand{\ee}{\end{equation}}
\newcommand{\bea}{\begin{eqnarray}}
\newcommand{\eea}{\end{eqnarray}}
\newcommand{\ba}{\begin{array}}
\newcommand{\ea}{\end{array}}
\newcommand{\bfi}[1]{\begin{figure}[#1]}
\newcommand{\efi}{\end{figure}}
\newcommand{\bpi}[2]{\begin{picture}(#1,#2)}
\newcommand{\epi}{\end{picture}}

\newcommand{\dis}{\displaystyle}
\renewcommand{\a}{\alpha}
\renewcommand{\b}{\beta}

\renewcommand{\d}{\delta}
\newcommand{\e}{\varepsilon}
\newcommand{\f}{\phi}
\newcommand{\g}{\gamma}

\newcommand{\m}{\mu}
\newcommand{\n}{\nu}

\newcommand{\p}{\pi}
\renewcommand{\r}{\rho}
\newcommand{\s}{\sigma}
\renewcommand{\t}{\tau}
\renewcommand{\th}{\theta}

\newcommand{\D}{\Delta}

\newcommand{\G}{\Gamma}
\renewcommand{\L}{\Lambda}

\renewcommand{\S}{\Sigma}

\newcommand{\phr}[1]{{\em Phys.\ Rev.\ }{\bf #1}}
\newcommand{\phrl}[1]{{\em Phys.\ Rev.\ Lett.\ }{\bf #1}}
\newcommand{\nph}[1]{{\em Nucl.\ Phys.\ }{\bf #1}}
\newcommand{\phl}[1]{{\em Phys.\ Lett.\ }{\bf #1}}
\newcommand{\zph}[1]{{\em Z.\ Phys.\ }{\bf #1}}

\newcommand{\jcph}[1]{{\em Jour.\ Comp. \ Phys.\ }{\bf #1}}

\input{feynman.tex}
\input{epsf.sty}

\begin{document}
\noindent
\hspace*{13.5cm}
UG--FT--65/96 \\
\hspace*{13.2cm}
hep-ph/9609259 \\
\hspace*{13.8cm}
August 1996
%
\begin{center}  \begin{Large} \begin{bf}
Bounds on the $Z\g\g$ couplings from HERA
 \footnote{Contribution to the Proceedings of the Workshop 
      ``Future Physics at HERA'', DESY, Hamburg, 1995-96.}\\
  \end{bf}  \end{Large}
  \vspace*{5mm}
  \begin{large}
F. Cornet$^a$, R. Graciani$^b$, J.I. Illana$^a$  
  \end{large}
\end{center}
$^a$ Depto. de F{\'\i}sica Te\'orica y del Cosmos,  
     Universidad de Granada, E-18071 Granada, Spain\\
$^b$ Depto. de F{\'\i}sica Te\'orica,
     Universidad Aut\'onoma de Madrid, E-28049 Cantoblanco, Spain \\
\begin{quotation}
\noindent
{\bf Abstract:}
The possibility of testing trilinear neutral gauge boson couplings
in radiative neutral current scattering at HERA is analyzed using a Monte
Carlo program that includes the Standard Model at tree level and the 
anomalous vertices. Acceptance and isolation cuts are 
applied as well as optimized cuts to enhance the signal from new physics.
The bounds on $Z\g\g$ couplings that can be achieved are not so
stringent as present bounds, even for high luminosities, but probe
a different kinematical region almost unsensitive to form
factors.
\end{quotation}
%

\section{Introduction}

The precision data collected to date have confirmed the
Standard Model to be a good description of physics below
the electroweak scale \cite{Schaile}. 
Despite of its great success, there are many reasons to believe 
that some kind of new physics must exist. On the other hand, the
non-abelian structure of the gauge  
boson self-couplings is still poorly tested and one of the most sensitive 
probes for new physics is provided by the trilinear gauge boson couplings 
(TGC) \cite{TGC}.

Many studies have been devoted to the $WW\g$ and $WWZ$ couplings.
At hadron colliders and $e^+e^-$ colliders, the present bounds
(Tevatron \cite{Errede}) and prospects (LHC, LEP2 and
NLC \cite{TGC,LEP2}) are mostly based on diboson production ($WW$, 
$W\g$ and $WZ$).
In $ep$ collisions, HERA could provide 
further information
analyzing single $W$ production ($ep\to eWX$ \cite{ABZ})  
and radiative charged current scattering 
($ep\to\n\g X$ \cite{hubert}). There is also some
literature on $WW\g$ couplings in $W$-pair production at future
very high energy photon colliders (bremsstrahlung photons in peripheral 
heavy ion collisions \cite{HIC} and Compton backscattered laser
beams \cite{gg}). 

Only recently, attention has been paid to the $Z\g Z$, $Z\g\g$ and
$ZZZ$ couplings. There is a detailed analysis of $Z\g V$
couplings ($V=\g,Z$) for hadron colliders in \cite{BB}. 
CDF \cite{CDF} and D\O\ \cite{D0}  have obtained bounds on the 
$Z\g Z$ and $Z\g\g$ anomalous couplings, while L3 has studied
only the first ones \cite{L3}. Studies on the sensitivities to
these vertices in future $e^+e^-$ colliders,
LEP2 \cite{LEP2} and NLC \cite{Boudjema}, have been performed during
the last years.
Some proposals have been made to probe these neutral boson gauge
couplings at future photon colliders in $e\g\to Ze$ \cite{eg}.

In this work we study the prospects for measuring the
TGC in the process $ep\to
e\g X$. In particular, we will concentrate on the $Z\g\g$ couplings,
which can be more stringently bounded than the $Z\g Z$ ones 
for this process.

In Section 2, we present the TGC. The next section deals with the
different contributions to the process $ep\to e\g X$ and the cuts
and methods we have employed 
in our analysis. Section 4 contains our results
for the Standard Model total cross section and distributions and
the estimates of the sensitivity of these quantities to the
presence of anomalous couplings. Finally, in the last section we
present our conclusions.

\section{Phenomenological parametrization of the neutral TGC}

A convenient way to study deviations from the standard model predictions
consists of considering the most general lagrangian compatible with
Lorentz invariance, the electromagnetic U(1) gauge symmetry, and
other possible gauge symmetries. 

For the trilinear $Z\g V$ couplings ($V=\g,Z)$ the most general vertex
function invariant under Lorentz and electromagnetic gauge transformations
can be described in terms of four independent dimensionless form
factors \cite{hagiwara}, denoted by $h^V_i$, i=1,2,3,4:
\bea
\G^{\a\b\m}_{Z\g V} (q_1,q_2,p)=\frac{f(V)}{M^2_Z}
\{
 h^V_1 (q^\m_2 g^{\a\b} - q^\a_2 g^{\m\b}) 
+\frac{h^V_2}{M^2_Z} p^\a (p\cdot q_2g^{\m\b}-q^\m_2 p^\b) 
\nonumber \\ 
+h^V_3 \e^{\m\a\b\r}q_{2_\r}
+\frac{h^V_4}{M^2_Z}p^\a\e^{\m\b\r\s}p_\r q_{2_\s}
\}. \hspace{3cm}
\label{vertex}
\eea
Terms proportional to $p^\m$, $q^\a_1$ and $q^\b_2$ are omitted as long as
the scalar components of all three vector bosons can be neglected
(whenever they couple to almost massless fermions) or they are zero 
(on-shell condition for $Z$ or U(1) gauge boson character of the photon).
The overall factor, $f(V)$, is $p^2-q^2_1$ for $Z\g Z$ or $p^2$ for $Z\g\g$
and is a result of Bose symmetry and electromagnetic gauge invariance.  
These latter constraints reduce the familiar seven form factors
of the most general $WWV$ vertex to only these four for the 
$Z\g V$ vertex. There still remains a global factor that can be fixed,
without loss of generality, to $g_{Z\g Z}=g_{Z\g\g}=e$. Combinations
of $h^V_3 (h^V_1)$ and $h^V_4 (h^V_2)$ correspond to electric
(magnetic) dipole and magnetic (electric) quadrupole transition
moments in the static limit.

All the terms are $C$-odd. The terms proportional to $h^V_1$ and $h^V_2$ 
are $CP$-odd while the other two are $CP$-even. All the form factors
are zero at tree level in the Standard Model. At the one-loop level,
only the $CP$-conserving $h^V_3$ and $h^V_4$ are nonzero \cite{barroso}
but too small (${\cal O}(\a/\p$)) to lead to any observable
effect at any present or planned experiment. However, larger effects 
might appear in theories or models beyond the Standard Model,
for instance when the gauge bosons are composite objects
\cite{composite}. 

This is a purely phenomenological, model independent parametrization. 
Tree-level unitarity restricts the $Z\g V$ to the Standard Model
values at asympotically high energies \cite{unitarity}. This
implies that the couplings $h^V_i$ have to be described by form factors 
$h^V_i(q^2_1,q^2_2,p^2)$ which vanish when $q^2_1$, $q^2_2$ or $p^2$
become large. In hadron colliders, large values of $p^2=\hat{s}$
come into play and the energy dependence has to be taken into
account, including unknown dumping factors \cite{BB}.
A scale dependence appears as an additional parameter (the scale
of new physics, $\L$). Alternatively,
one could introduce a set of operators invariant under SU(2)$\times$U(1)
involving the gauge bosons and/or additional would-be-Goldstone bosons
and the physical Higgs. Depending on the new physics dynamics,
operators with dimension $d$ could be generated at the scale $\L$, 
with a strength which is generally suppressed by factors like
$(M_W/\L)^{d-4}$ or $(\sqrt{s}/\L)^{d-4}$ \cite{NPscale}.  
It can be shown that $h^V_1$ and $h^V_3$ receive contributions from
operators of dimension $\ge 6$ and $h^V_2$ and $h^V_4$ from
operators of dimension $\ge 8$.
Unlike hadron colliders, in $ep\to e\g X$ at HERA energies, we can ignore 
the dependence of the form factors on the scale. On the other
hand, the anomalous couplings are tested in a different kinematical region,
which makes their study in this process complementary to the ones
performed at hadron and lepton colliders.

\section{The process $ep\to e\g X$}

The process under study is $ep\to e\g X$, which is described in the
parton model by the radiative neutral current electron-quark and 
electron-antiquark scattering, 
\be
\label{process}
e^- \ \stackrel{(-)}{q} \to e^- \ \stackrel{(-)}{q} \ \g .
\ee

There are eight Feynman diagrams contributing to this process in
the Standard Model and three additional ones if one includes anomalous vertices:
one extra diagram for the $Z\g Z$ vertex and two for the $Z\g\g$
vertex (Fig. \ref{feyndiag}).

\bfi{htb}
\begin{center}
\bigphotons
\bpi{35000}{21000}
%
%
\put(4000,8000){(a)}
\put(200,17000){\vector(1,0){1300}}
\put(1500,17000){\vector(1,0){3900}}
\put(5400,17000){\line(1,0){2600}}
\drawline\photon[\S\REG](2800,17000)[5]
\put(200,\pbacky){\vector(1,0){1300}}
\put(1500,\pbacky){\vector(1,0){2600}}
\put(4100,\pbacky){\vector(1,0){2600}}
\put(6700,\pbacky){\line(1,0){1300}}
\put(0,13000){$q$}
\put(8200,13000){$q$}
\put(3300,\pmidy){$\g,Z$}
\drawline\photon[\SE\FLIPPED](4900,\pbacky)[4]
\put(0,18000){$e$}
\put(8200,18000){$e$}
\put(8200,\pbacky){$\g$}
%
%
\put(13000,8000){(b)}
\put(9500,17000){\vector(1,0){1300}}
\put(10800,17000){\vector(1,0){2600}}
\put(13400,17000){\vector(1,0){2600}}
\put(16000,17000){\line(1,0){1300}}
\drawline\photon[\S\REG](12100,17000)[5]
\put(9500,\pbacky){\vector(1,0){1300}}
\put(10800,\pbacky){\vector(1,0){3900}}
\put(14700,\pbacky){\line(1,0){2600}}
\drawline\photon[\NE\FLIPPED](14200,17000)[4]
%
%
\put(22000,8000){(c)}
\put(18500,17000){\vector(1,0){3250}}
\put(21750,17000){\vector(1,0){3250}}
\put(25000,17000){\line(1,0){1300}}
\drawline\photon[\S\REG](23700,17000)[5]
\put(18500,\pbacky){\vector(1,0){1300}}
\put(19800,\pbacky){\vector(1,0){2600}}
\put(22400,\pbacky){\vector(1,0){2600}}
\put(25000,\pbacky){\line(1,0){1300}}
\drawline\photon[\SE\FLIPPED](21100,\pbacky)[4]
%
%
\put(31000,8000){(d)}
\put(27500,17000){\vector(1,0){1300}}
\put(28800,17000){\vector(1,0){2600}}
\put(31400,17000){\vector(1,0){2600}}
\put(34000,17000){\line(1,0){1300}}
\drawline\photon[\S\REG](32700,17000)[5]
\put(27500,\pbacky){\vector(1,0){3250}}
\put(30750,\pbacky){\vector(1,0){3250}}
\put(33900,\pbacky){\line(1,0){1300}}
\drawline\photon[\NE\FLIPPED](30100,17000)[4]
%
%
\put(17800,0){(e)}
\put(17100,5500){$\g,Z$}
\put(17100,3000){$\g,Z$}
\put(14000,7000){\vector(1,0){1300}}
\put(15300,7000){\vector(1,0){3900}}
\put(19200,7000){\line(1,0){2600}}
\drawline\photon[\S\REG](16600,7000)[5]
\put(16750,\pmidy){\circle*{500}}
\put(14000,\pbacky){\vector(1,0){1300}}
\put(15300,\pbacky){\vector(1,0){3900}}
\put(19200,\pbacky){\line(1,0){2600}}
\drawline\photon[\E\REG](16750,\pmidy)[5]
\put(22300,\pbacky){$\g$}
\epi
\end{center}
\caption{\it Feynman diagrams for the process $e^- q \to e^- q \g$.
	\label{feyndiag}}
\efi

Diagrams with $\g$ exchanged in the t-channel are dominant. Nevertheless,
we consider the whole set of diagrams in the calculation.
On the other side, u-channel fermion exchange poles appear, in the limit
of massless quarks and electrons (diagrams (c) and (d)). 
Since the anomalous diagrams (e) do not present such infrared or
collinear singularities, it seems appropriate to avoid almost
on-shell photons exchanged and fermion poles by cutting the
transverse momenta of the final fermions (electron and jet) to
enhance the signal from anomalous vertices. 
Due to the suppression factor coming from $Z$ propagator, the 
anomalous diagrams are more sensitive to $Z\g\g$ than to $Z\g Z$ vertices.
In the following we will focus our attention on the former.

The basic variables of the parton level process are five. A 
suitable choice is: $E_\g$ (energy of the final photon), 
$\cos\th_\g$, $\cos\th_{q'}$ (cosines of the polar angles of the 
photon and the scattered quark defined with respect to the proton direction),
$\f$ (the angle between the transverse momenta of the photon and the
scattered quark in a plane perpendicular to the beam), and a
trivial azimuthal angle that is integrated out (unpolarized beams).
All the variables are referred to the laboratory frame. One needs
an extra variable, the Bjorken-x, to connect the partonic process
with the $ep$ process. The phase space integration over these six
variables is carried out by {\tt VEGAS} \cite{VEGAS} and has been 
cross-checked with the {\tt RAMBO} subroutine \cite{RAMBO}.


We adopt two kinds of event cuts to constrain conveniently
the phase space:

\begin{itemize}

\item 
{\em Acceptance and isolation} cuts. The former are to exclude 
phase space regions
which are not accessible to the detector, because of angular or
efficiency limitations:\footnote{The threshold for the transverse
momentum of the scattered quark ensures that its kinematics can be
described in terms of a jet.}
\bea
\label{cut1}
8^o < \theta_e,\ \theta_\g,\ \theta_{\rm jet} < 172^o; \nonumber\\
E_e, \ E_\g, \ p^{\rm q'}_{\rm T} > 10 \ {\rm GeV}.
\eea
The latter keep the final photon well separated
from both the final electron and the jet:
\bea
\label{cut2}
\cos \langle \g,e \rangle < 0.9; \nonumber\\
R > 1.5,
\eea
where $R\equiv\sqrt{\D\eta^2+\phi^2}$ is the separation between
the photon and the jet in the rapidity-azimuthal plane, and $\langle \g,e \rangle$ is the angle between the photon and the scattered electron.

\item
Cuts for {\em intrinsic background suppression}. They consist of 
strengthening some of the 
previous cuts or adding new ones to enhance the signal of the anomalous
diagrams against the Standard Model background.

\end{itemize}

We have developed a Monte Carlo program for the simulation of the
process $ep\to e\g X$ where $X$ is the remnant of the proton plus one jet
formed by the scattered quark of the subprocess (\ref{process}). It 
includes the Standard Model helicitity amplitudes computed using the {\tt HELAS} subroutines \cite{HELAS}. We added new code to account for the
anomalous diagrams. The squares of these anomalous amplitudes have been 
cross-checked with their analytical expressions computed using {\tt FORM}
\cite{FORM}. For the parton distribution functions, 
we employ both the set 1 of Duke-Owens' parametrizations \cite{DO}
and the modified MRS(A) parametrizations \cite{MRS}, with the scale chosen to
be the hadronic momentum transfer.

As inputs, we use the beam energies $E_e=30$ GeV and $E_p=820$ GeV, 
the $Z$ mass $M_Z=91.187$ GeV, the weak angle $\sin^2_W=0.2315$ 
\cite{PDB} and the fine structure constant $\a=1/128$. A more correct choice
would be the running fine structure constant with $Q^2$ as the argument.
However, as we are interested in large $Q^2$ events, the value $\a(M^2_Z)$
is accurate enough for our purposes. We consider only 
the first and second generations of quarks, assumed to be massless.

We start by applying the cuts (\ref{cut1}) and (\ref{cut2})
and examining the contribution to a set of observables of the
Standard Model and the anomalous diagrams, separately. Next, we 
select one observable such that, when a cut on it is performed,
only Standard Model events are mostly eliminated. The procedure
is repeated with this new cut built in. After several runs, adding
new cuts, the ratio standard/anomalous cross sections is reduced
and hence the sensitivity to anomalous couplings is improved. 

\section{Results}

\subsection{Observables}

The total cross section of $ep\to e\g X$ can be written as
\be
\s=\s_{{\rm SM}} + \sum_{i} \t_i \cdot h^\g_i + \sum_{i}\s_i\cdot (h^\g_i)^2
    + \s_{12} \cdot h^\g_1 h^\g_2 + \s_{34} \cdot h^\g_3 h^\g_4.
\ee

\bfi{htb}
\setlength{\unitlength}{1cm}
\bpi{8}{7}
\epsfxsize=11cm
\put(-1,-4){\epsfbox{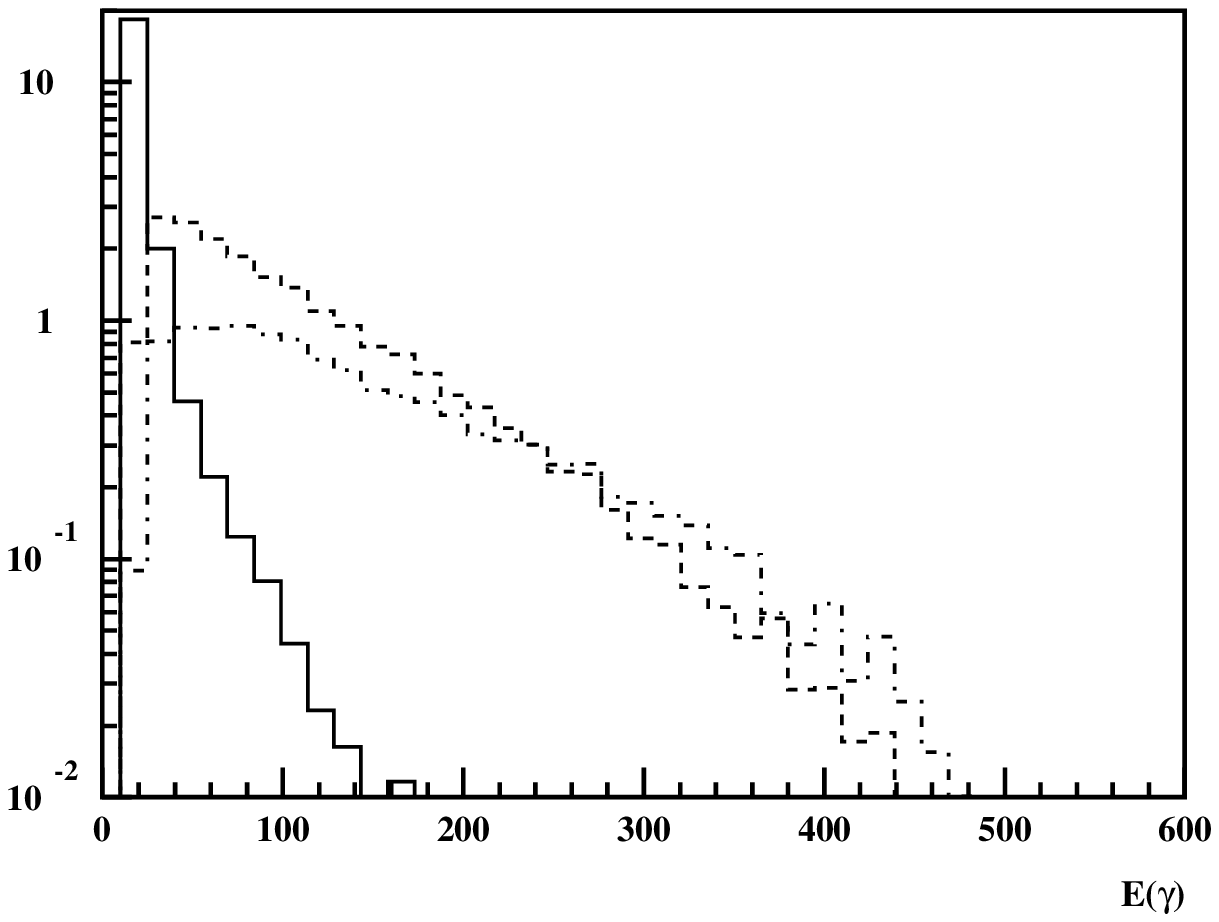}}
\epi
\bpi{8}{7}
\epsfxsize=11cm
\put(0.,-4){\epsfbox{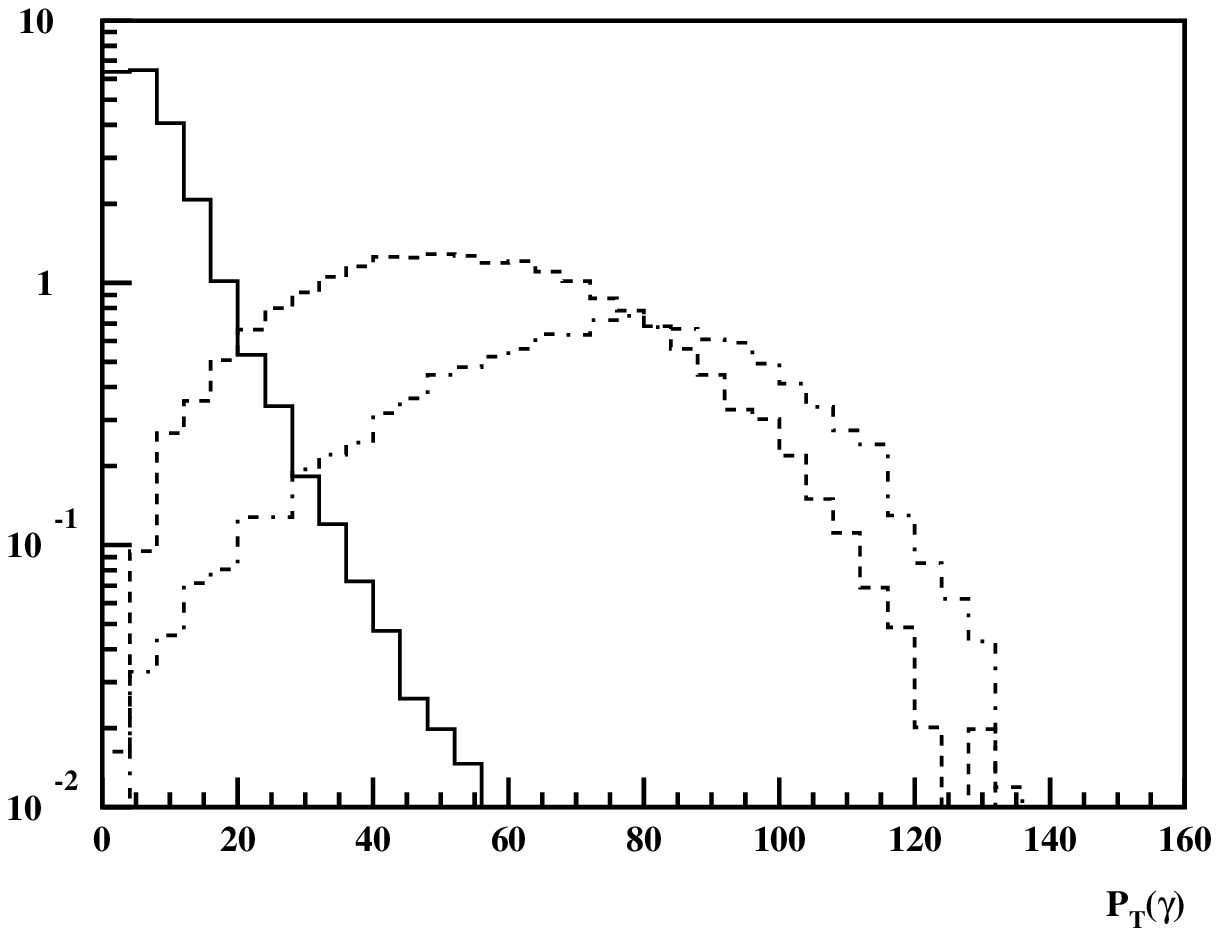}}
\epi

\bpi{8}{6}
\epsfxsize=11cm
\put(-1,-5){\epsfbox{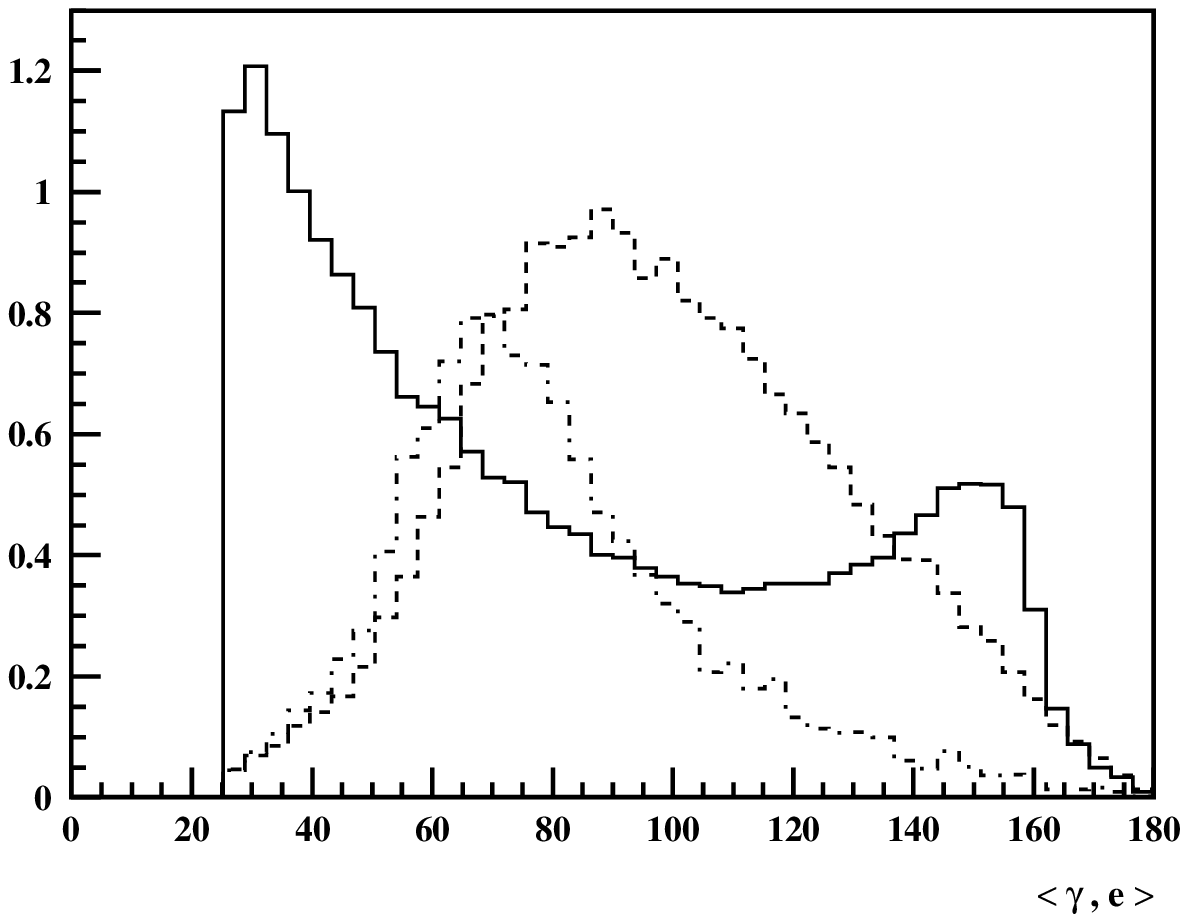}}
\epi
\bpi{8}{6}
\epsfxsize=11cm
\put(0.,-5){\epsfbox{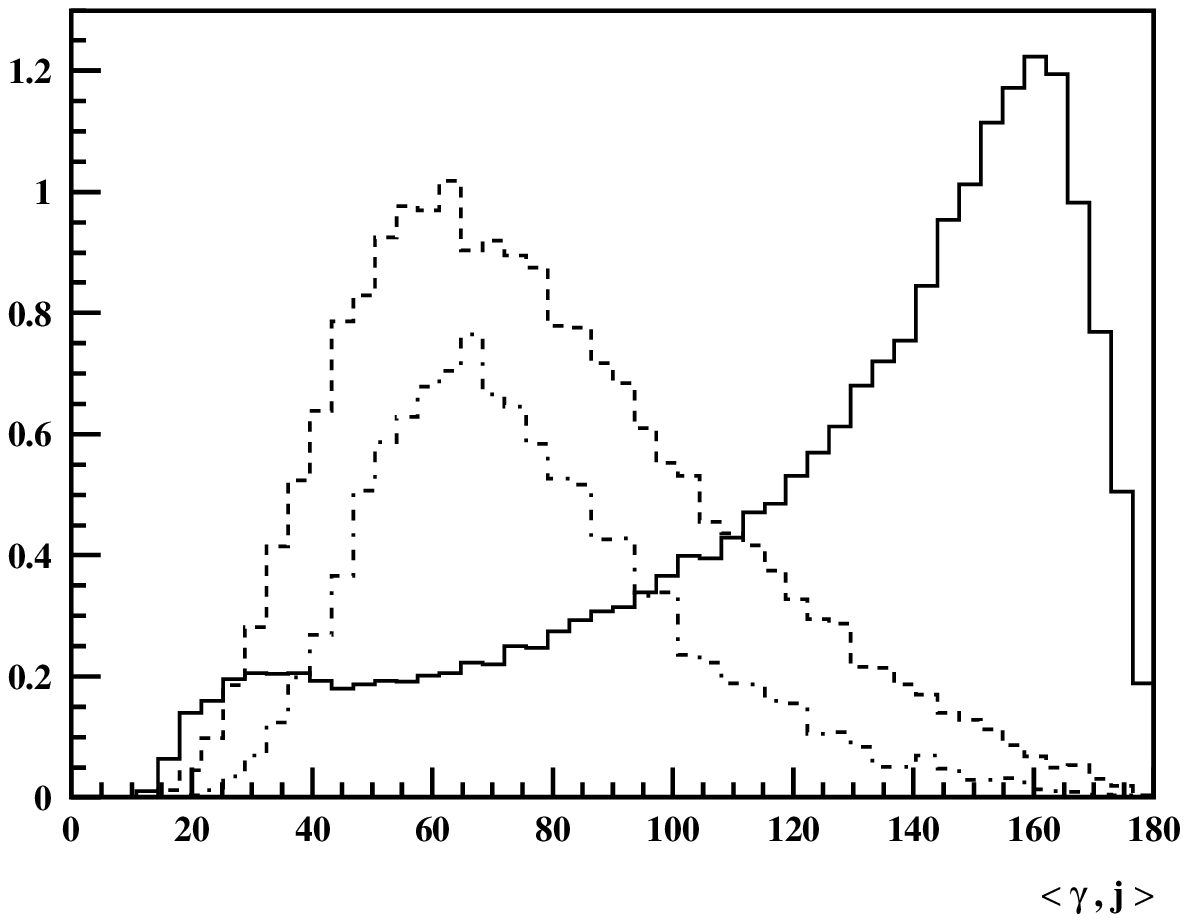}}
\epi

\bpi{8}{6}
\epsfxsize=11cm
\put(-1,-5){\epsfbox{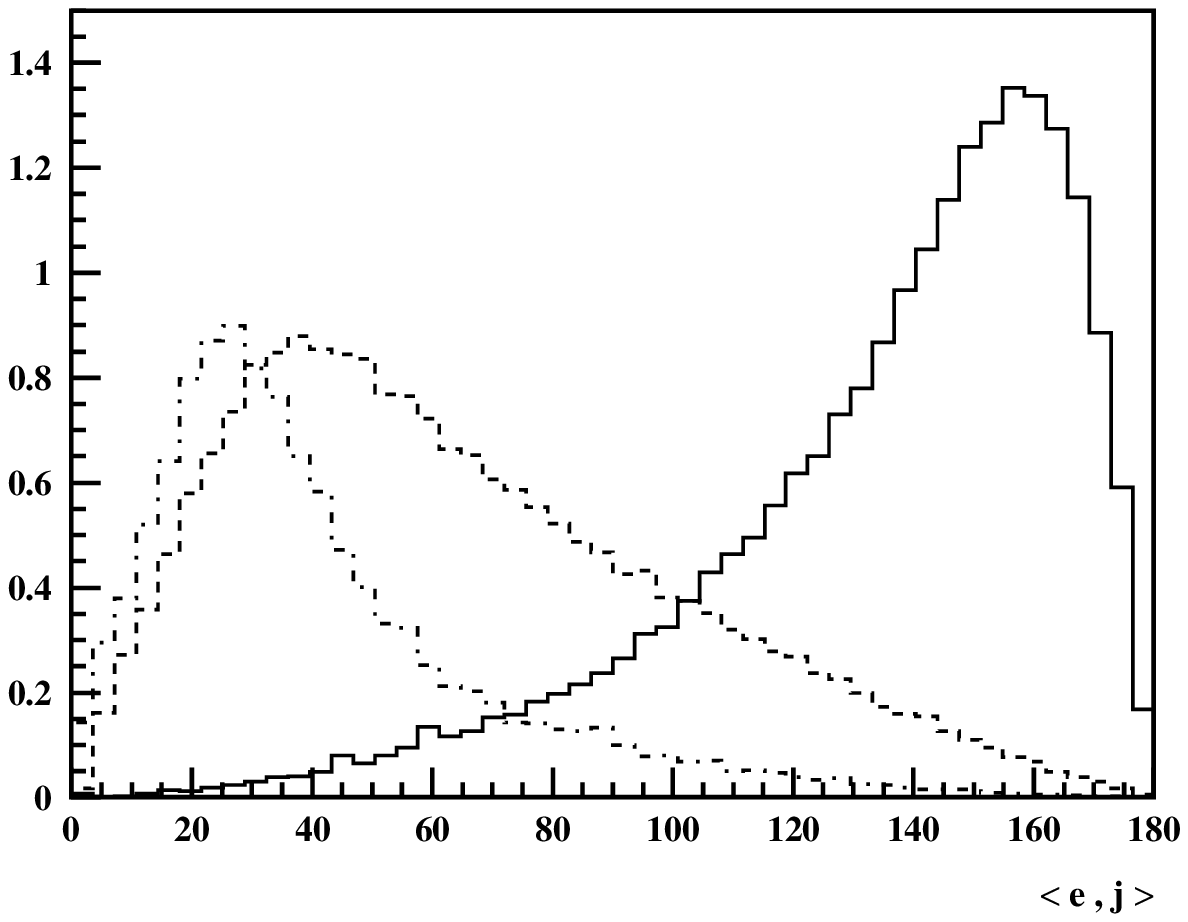}}
\epi
\bpi{8}{7}
\epsfxsize=11cm
\put(0.,-5){\epsfbox{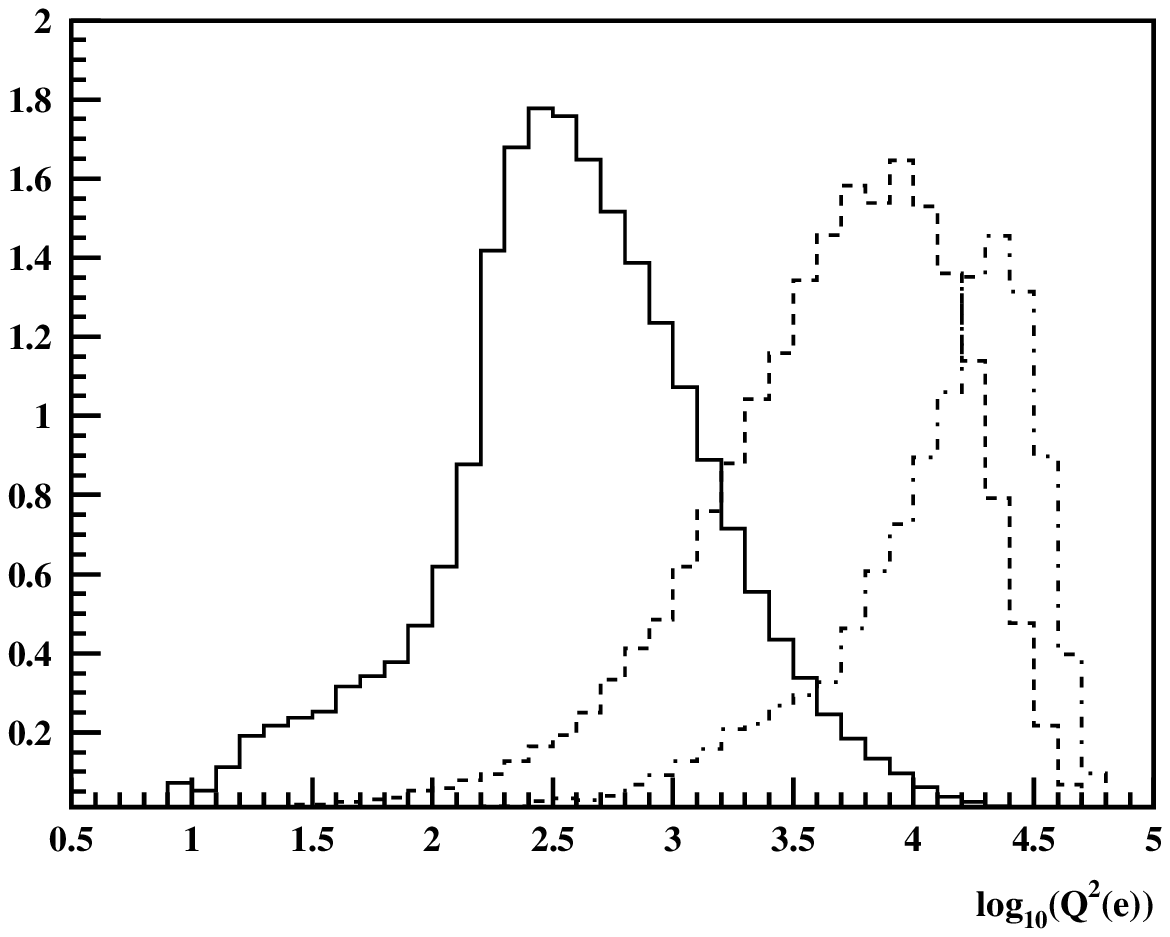}}
\epi
\caption{\it Differential cross sections (pb) for the process $ep\to e\g X$ at
HERA, with only acceptance and isolation cuts.
The solid line is the Standard Model contribution and the dash (dot-dash) line 
correspond to 10000 times the $\s_1$ ($\s_2$) anomalous contributions.\label{A}}
\efi

The forthcoming results are obtained using the MRS'95
pa\-ra\-me\-tri\-za\-tion of the parton densities\footnote{The values 
change $\sim 10$\% when using the (old) Duke-Owens' structure functions.}
\cite{MRS}.
The linear terms of the $P$-violating couplings $h^\g_3$
and $h^\g_4$ are negligible, as they mostly arise from the interference of
standard model diagrams with photon exchange ($P$-even) and anomalous
$P$-odd diagrams ($\t_3\simeq \t_4\simeq 0$). Moreover, anomalous diagrams with
different $P$ do not interfere either. On the other hand, the quadratic terms 
proportional to $(h^\g_1)^2$ and $(h^\g_3)^2$ have identical expressions, and 
the same for $h^\g_2$ and $h^\g_4$ ($\s_1=\s_3$, $\s_2=\s_4$). Only the 
linear terms make their bounds different. The interference terms $\s_{12}$ 
and $\s_{34}$ are also identical.

\bfi{htb}
\setlength{\unitlength}{1cm}
\bpi{8}{7}
\epsfxsize=11cm
\put(-1,-4){\epsfbox{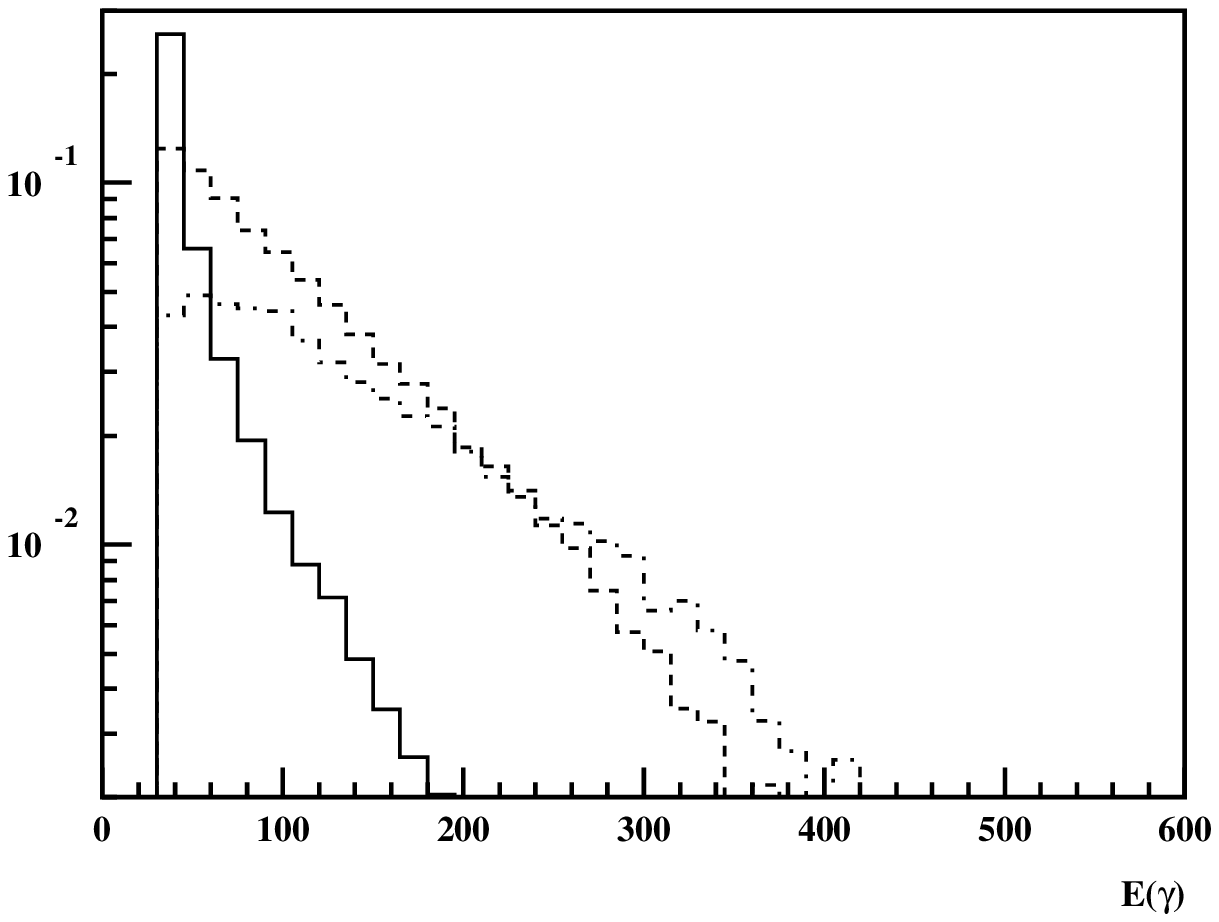}}
\epi
\bpi{8}{7}
\epsfxsize=11cm
\put(0.,-4){\epsfbox{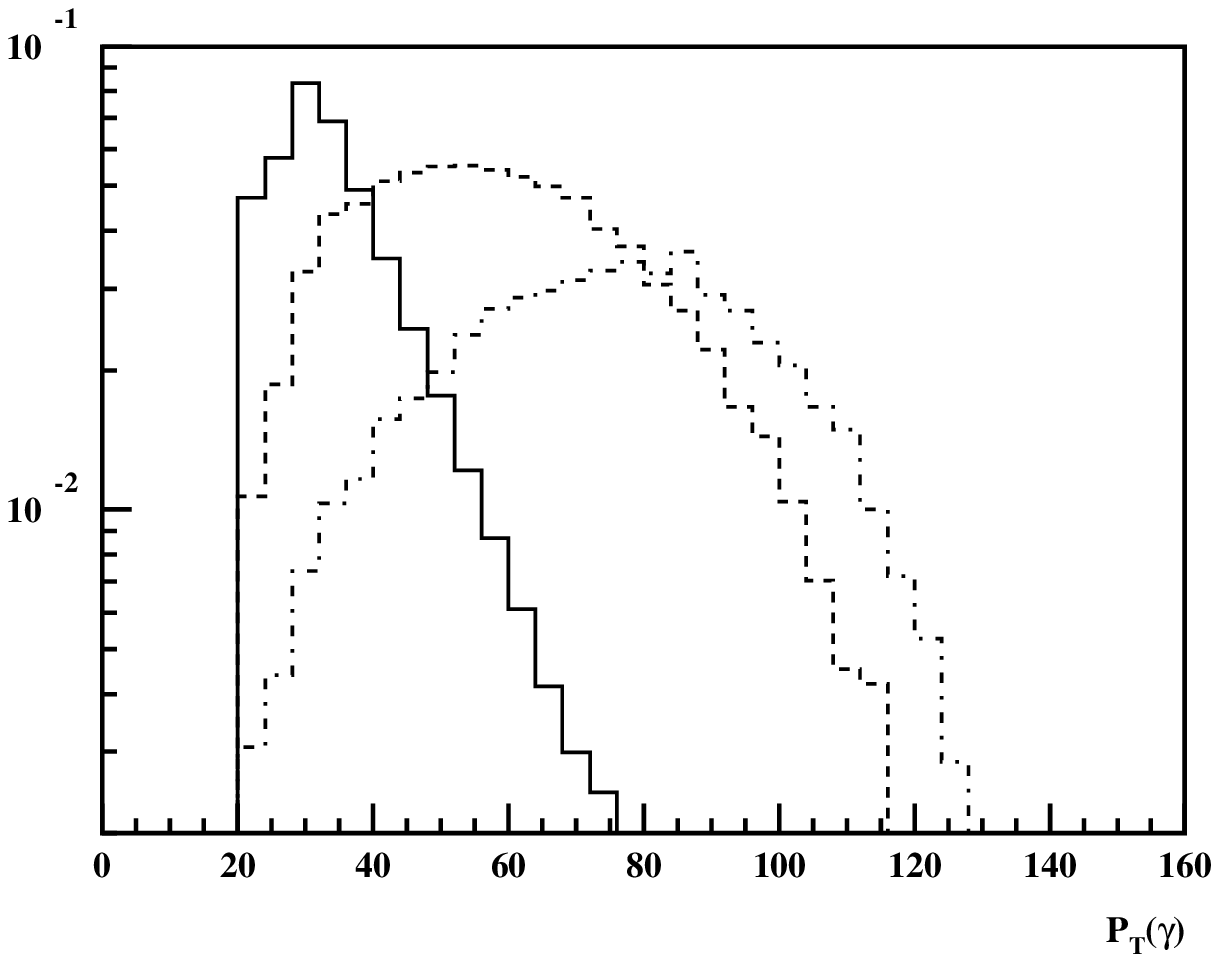}}
\epi

\bpi{8}{6}
\epsfxsize=11cm
\put(-1,-5){\epsfbox{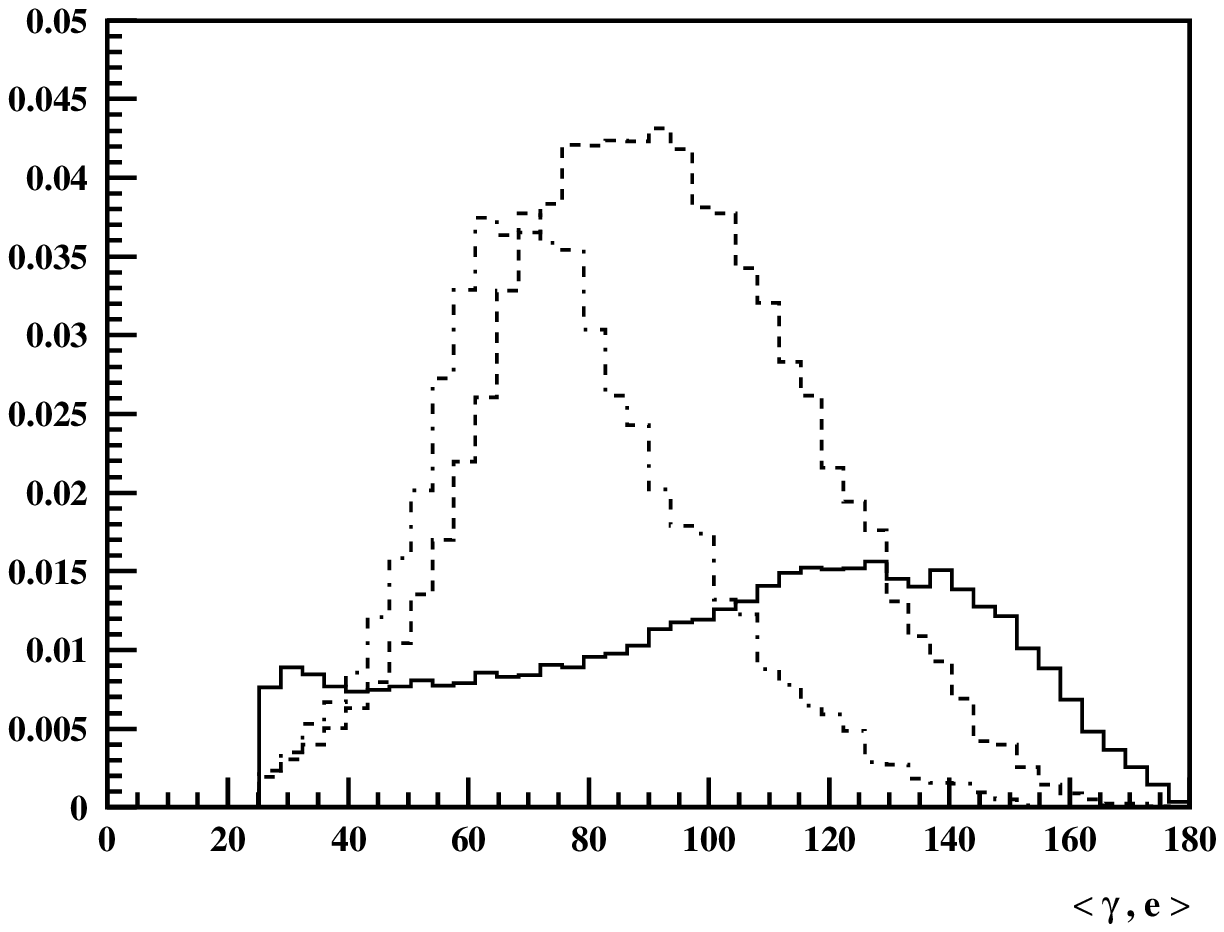}}
\epi
\bpi{8}{6}
\epsfxsize=11cm
\put(0.,-5){\epsfbox{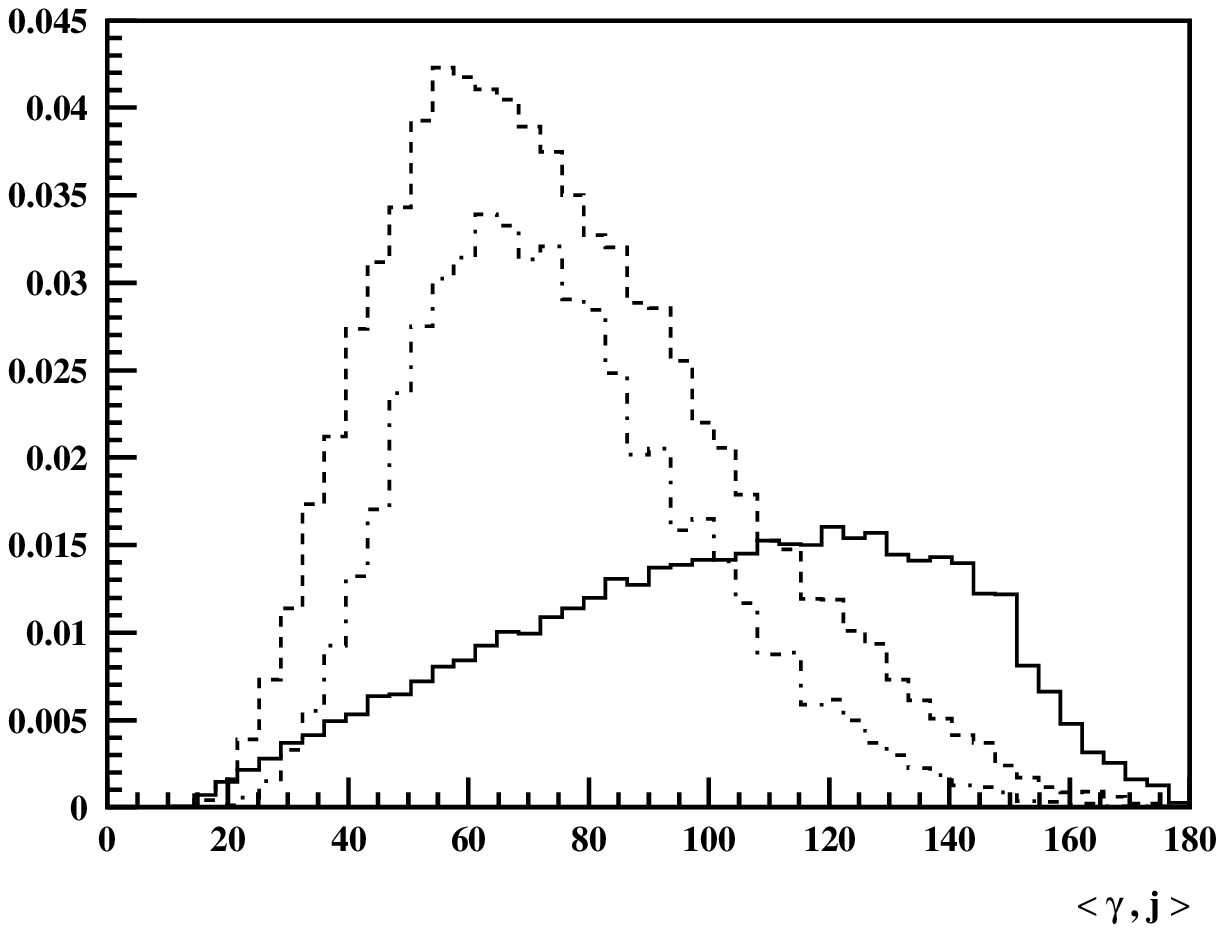}}
\epi

\bpi{8}{6}
\epsfxsize=11cm
\put(-1,-5){\epsfbox{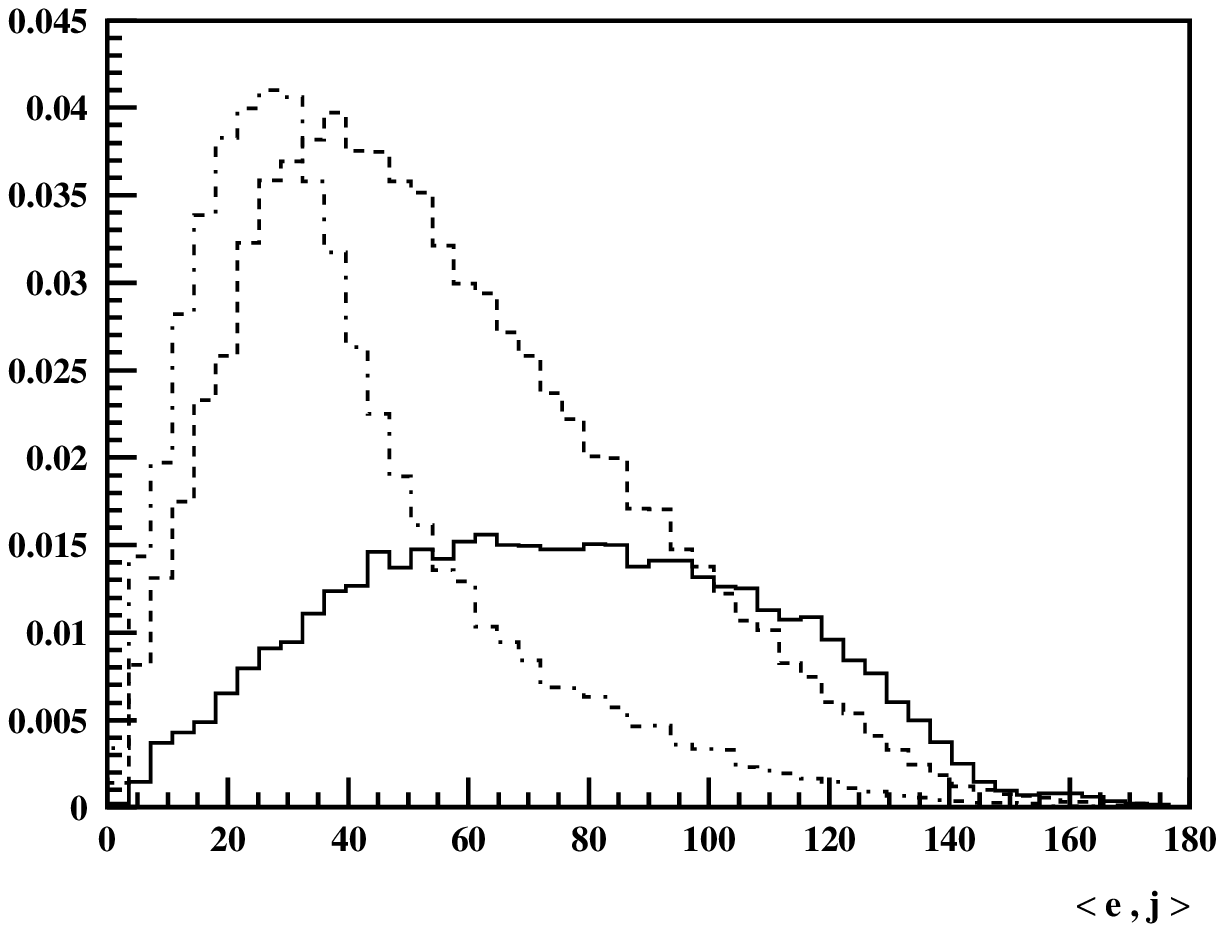}}
\epi
\bpi{8}{7}
\epsfxsize=11cm
\put(0.,-5){\epsfbox{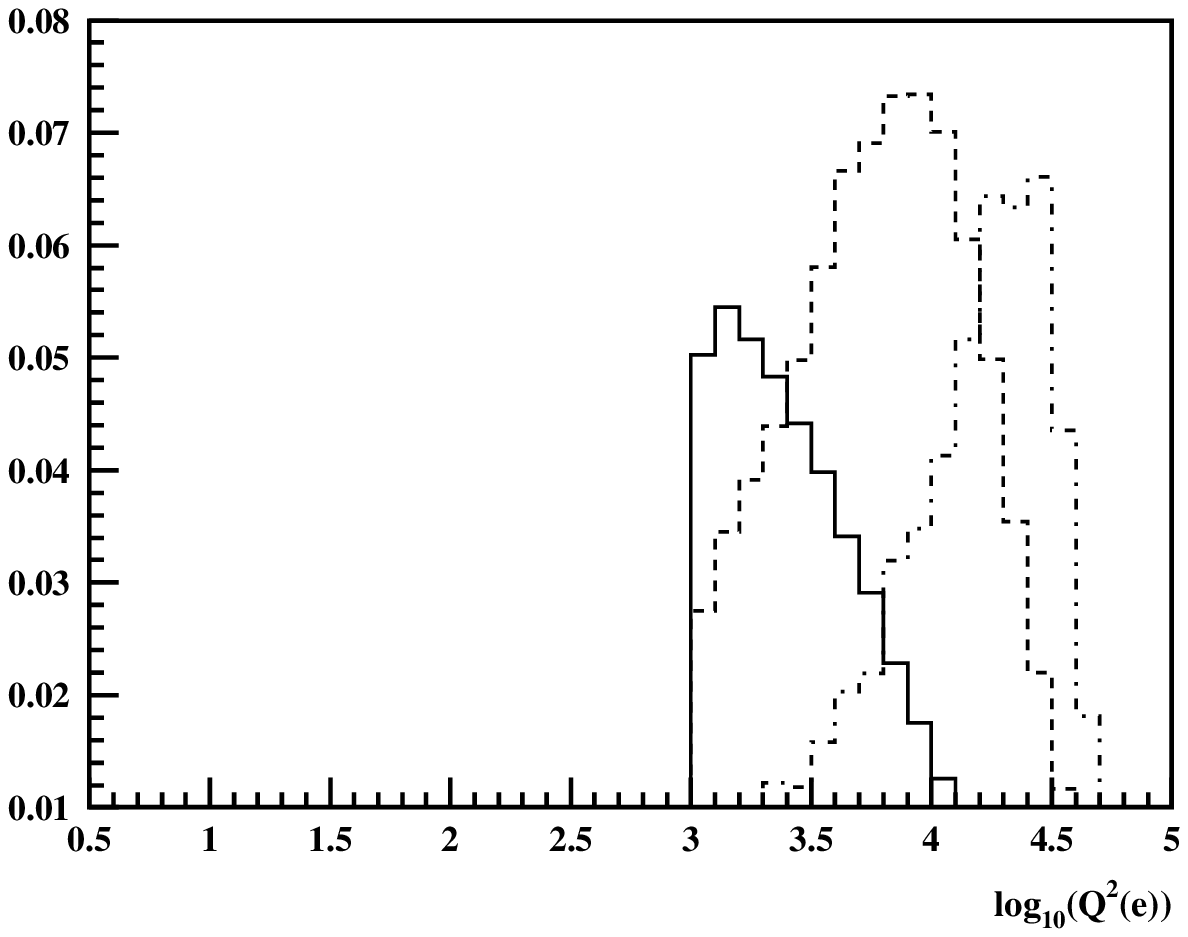}}
\epi
\caption{\it Differential cross sections (pb) for the process $ep\to e\g X$ at
HERA, after intrinsic background suppression.
The solid line is the Standard Model contribution and the dash (dot-dash) line correspond to 500 times the $\s_1$ ($\s_2$) anomalous contributions.\label{B}}
\efi

We have analyzed the distributions of more than twenty
observables in the laboratory frame, including the energies, transverse
momenta and angular distributions of the jet, the photon and the final 
electron, as well as their spatial, polar and azimuthal separations.
Also the bjorken-x, the leptonic and hadronic momenta transfer and other fractional energies are considered.

The process of intrinsic background suppression is illustrated
by comparing Figures \ref{A} and \ref{B}. For simplicity, only
the most interesting variables are shown: the energy $E(\g)$ and transverse
momentum $p_T(\g)$ of the photon; the angles between the photon and
the scattered electron $\langle \g,e \rangle$, the photon and the jet
$\langle \g,j \rangle$, and the scattered electron and the jet $\langle e,j
\rangle$; and the leptonic momentum transfer $Q^2(e)$.
In Fig.~\ref{A}, these variables
are plotted with only acceptance and isolation cuts 
implemented.
All of them share the property of disposing of a range
where any anomalous effect is negligible, whereas the contribution
to the total SM cross section is large. The set of cuts
listed below were added to reach eventually the distributions of
Fig.~\ref{B}:

\begin{itemize}

\item
The main contribution to the Standard Model cross section comes from
soft photons with very low transverse momentum. The following cuts
suppress a 97$\%$ of these events, without hardly affecting the
anomalous diagrams which, conversely, enfavour high energy photons:
\bea
E_\g > 30 \ {\rm GeV} \nonumber \\
p^\g_T > 20 \ {\rm GeV}
\label{cut3}
\eea

\item
Another remarkable feature of anomalous diagrams is the very different
typical momentum transfers. Let's concentrate on the leptonic momentum
transfer, $Q^2_e=-(p'_e-p_e)^2$. The phase space enhances high
$Q^2_e$, while the photon propagator of the Standard Model diagrams
prefer low values (above the threshold for electron detectability,
$Q^2_e>5.8$~GeV$^2$, with our required minimum energy and angle). On the
contrary, the anomalous diagrams have always a $Z$ propagator
which introduces a suppression factor of the order of $Q^2_e/M^2_Z$ and
makes irrelevant the $Q^2_e$ dependence, which is only determined by
the phase space. As a consequence, the following cut looks appropriate, 
\be
Q^2_e > 1000 \ {\rm GeV}^2
\label{cut4}
\ee

\end{itemize}

It is important to notice at this point why usual form factors for the
anomalous couplings can be neglected at HERA. For our process, these
form factors should be proportional to $1/(1+Q^2/\L^2)^n$. With the scale of 
new physics $\L=500$~GeV to 1~TeV, these factors can be taken to be one. This 
is not the case in lepton or hadron high energy colliders where the diboson production in the s-channel needs dumping factors $1/(1+\hat{s}/\L^2)^n$.    

The total cross section for the Standard Model with acceptance and isolation
cuts is $\s_{\rm SM}=21.38$~pb and is reduced to 0.37~pb when all the cuts are applied, while the quadratic contributions only change from 
$\s_1=2\times10^{-3}$~pb, $\s_2=1.12\times10^{-3}$~pb to 
$\s_1=1.58\times10^{-3}$~pb, $\s_2=1.05\times10^{-3}$~pb. The linear
terms are of importance and change from $\t_1=1.18\times10^{-2}$~pb, $\t_2=1.27\times10^{-3}$~pb to $\t_1=7.13\times10^{-3}$~pb, $\t_2=1.26\times10^{-3}$~pb. Finally, the interference term $\s_{12}=1.87\times10^{-3}$~pb changes to $\s_{12}=1.71\times10^{-3}$~pb.


The typical Standard Model events consist of soft and low-$p_T$ photons
mostly backwards, tending to go in the same direction of the scattered
electrons (part of them are emitted by the hadronic
current in the forward direction), close to the required angular separation ($\sim 30^o$). The low-$p_T$ jet goes opposite to both the photon and the scattered electron, also in the transverse plane.
On the contrary, the anomalous events have not so soft and high-$p_T$ photons,
concentrated in the forward region as it the case for the scattered electron 
and the jet.

\subsection{Sensitivity to anomalous couplings}

In order to estimate the sensitivity to anomalous couplings, we
consider the $\chi^2$ function.
One can define the $\chi^2$, which is related to the likelihood
function ${\cal L}$, as
\be
\label{chi2}
\chi^2\equiv-2\ln{\cal L}=
2 L \dis\left(\s^{th}-\s^{o}+\s^{o}
\ln\dis\frac{\s^{o}}{\s^{th}}\right)
\simeq L \dis\dis\frac{(\s^{th}-\s^{o})^2}{\s^{o}},
\ee
where $L=N^{th}/\s^{th}=N^o/\s^o$ is the integrated luminosity
and $N^{th}$ ($N^o$) is the number of theoretical (observed)
events. The last line of (\ref{chi2}) is a useful
and familiar approximation, only valid when $\mid \s^{th}-\s^o \mid/
\s^o \ll 1$. 
This function is a measure of the probability that statistical
fluctuations can make undistinguisable the observed and the predicted 
number of events, that is the Standard Model prediction. The well
known $\chi^2$-CL curve allows us to determine the corresponding
confidence level (CL).
We establish bounds on the anomalous couplings by fixing a
certain $\chi^2=\d^2$ and allowing for the $h^\g_i$
values to vary, $N^o=N^o(h^\g_i)$. The parameter $\d$ is often referred 
as the number of
standard deviations or `sigmas'. A $95\%$ CL corresponds to almost
two sigmas ($\d=1.96$). 
When $\s \simeq \s_{{\rm SM}} + (h^\g_i)^2 \s_i$ (case of the $CP$-odd
terms) and the anomalous contribution is small enough, the
upper limits present some useful, approximate scaling properties,
with the luminosity,
\be
h^\g_i (L')\simeq\sqrt[4]{\frac{L}{L'}} \ h^\g_i (L).
\ee

A brief comment on the interpretation of the results is in order.
As the cross section grows with $h^\g_i$, in the relevant range of
values, the $N^o$ upper limits can be regarded as the lowest number
of measured events that would discard the Standard Model, or the
largest values of $h^\g_i$ that could be bounded if no effect is
observed, with the given CL. This procedure approaches the 
method of upper limits for Poisson processes when the
number of events is large ($\gsim 10$).

\bfi{htb}
\setlength{\unitlength}{1cm}
\bpi{8}{8}
\epsfxsize=12cm
\put(3.35,4.245){+}
\put(-2.5,-1.5){\epsfbox{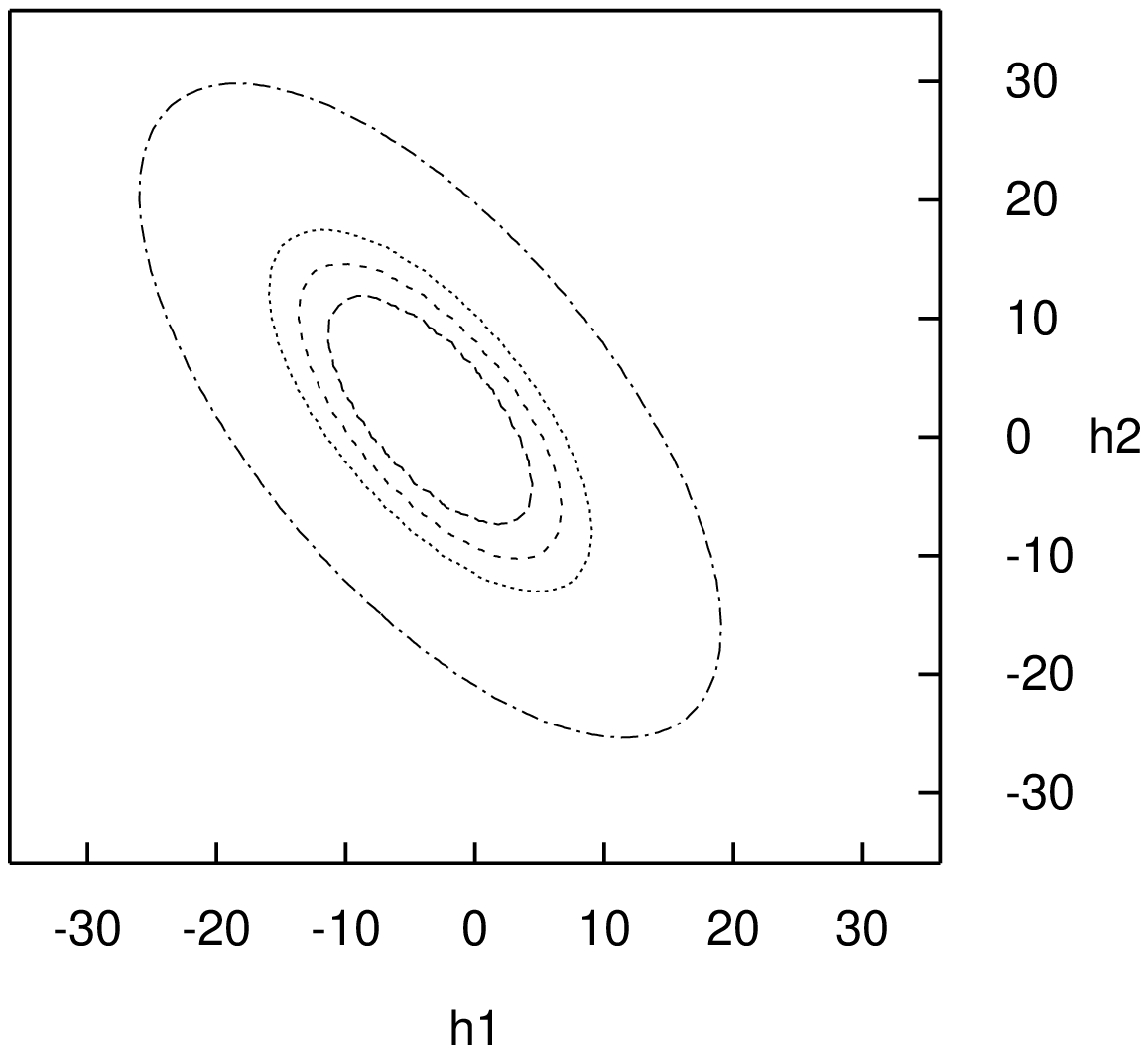}}
\epi
\bpi{8}{8}
\epsfxsize=12cm
\put(4.1,4.245){+}
\put(-1.75,-1.5){\epsfbox{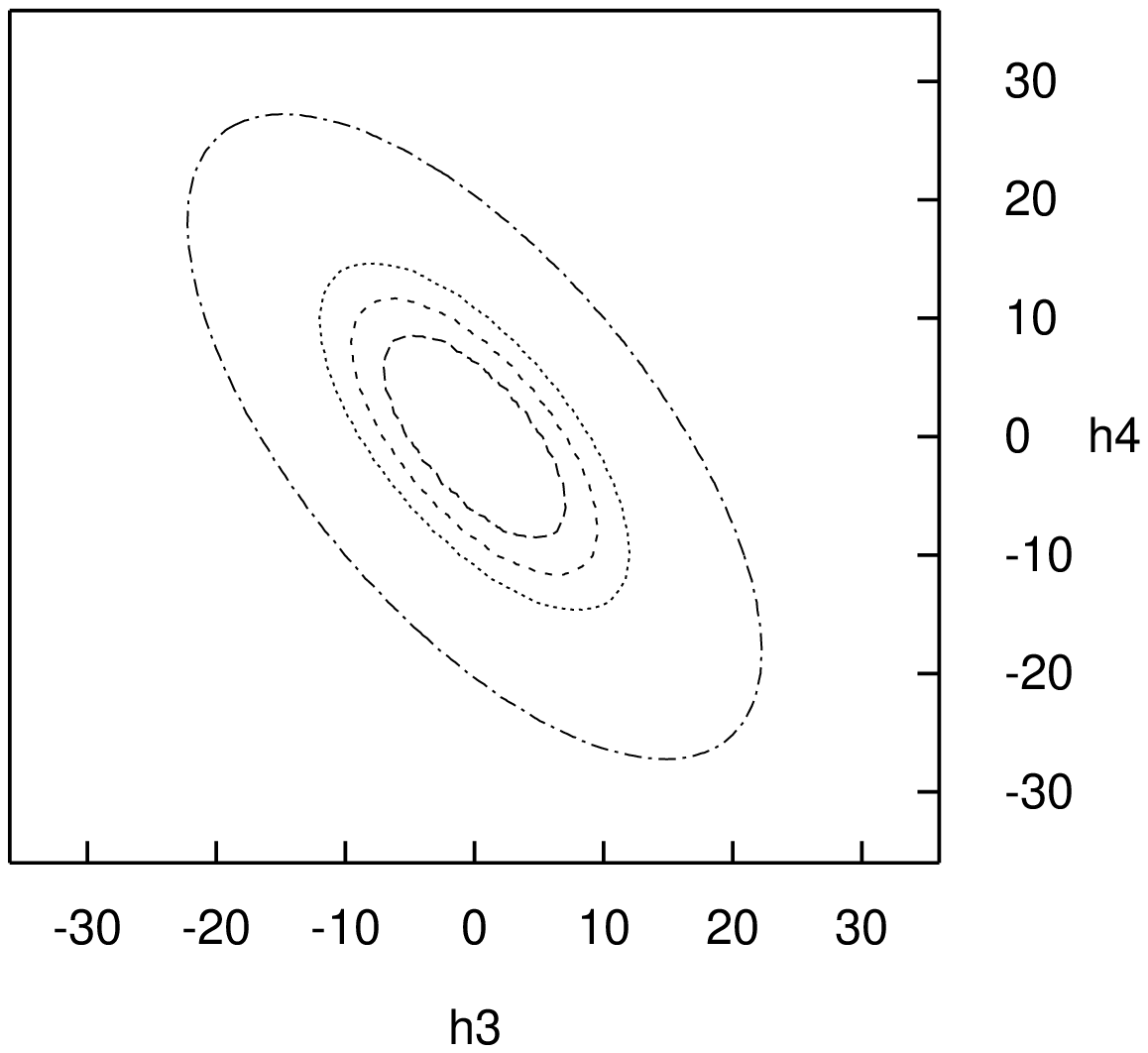}}
\epi
\caption{\it Limit contours for $Z\g\g$ couplings at HERA with an integrated luminosity of 10, 100, 250, 1000 pb$^{-1}$ and a 95\% CL.\label{contour}}
\efi

In Fig. \ref{contour} the sensitivities for different luminosities are shown.
Unfortunately, HERA cannot compete with Tevatron, whose best 
bounds, reported by the D\O\ collaboration \cite{D0}, are
\bea
|h^\g_1|, \ |h^\g_3| &<& 1.9 \ (3.1), \nonumber
\\
|h^\g_2|, \ |h^\g_4| &<& 0.5 \ (0.8).
\eea
For the first value it was assumed that only one anomalous coupling contributes
(`axial limits') and for the second there are two couplings contributing (`correlated limits'). Our results are summarized in Table \ref{table}.

\begin{table}
\begin{center}
\begin{tabular}{|c|r|r|r|r|r|r|r|r|}
\hline
HERA & \multicolumn{2}{c|}{10 pb$^{-1}$}  & \multicolumn{2}{c|}{100 pb$^{-1}$}
     & \multicolumn{2}{c|}{250 pb$^{-1}$} & \multicolumn{2}{c|}{1 fb$^{-1}$} \\
\hline \hline
$h^\g_1$ & -19.0 & 14.5 & -11.5 &  7.0 &  -9.5 &  5.5 &  -8.0 &  3.5 \\
         & -26.0 & 19.5 & -16.0 &  9.5 & -14.0 &  7.0 & -11.5 &  4.5 \\
\hline
$h^\g_2$ & -21.5 & 20.0 & -12.0 & 10.0 & - 9.5 &  8.0 &  -7.0 &  6.0 \\
         & -26.0 & 30.0 & -13.0 & 18.0 & -10.0 & 15.0 & - 7.5 & 12.0 \\
\hline
$h^\g_3$ & -17.0 & 17.0 &  -9.0 &  9.0 &  -7.5 &  7.5 &  -5.5 &  5.5 \\
         & -22.5 & 22.5 & -12.0 & 12.0 & -10.0 & 10.0 &  -7.0 &  7.0 \\
\hline 
$h^\g_4$ & -20.5 & 20.5 & -11.0 & 11.0 &  -8.5 &  8.5 &  -6.0 &  6.0 \\
         & -27.5 & 27.5 & -14.5 & 14.5 & -12.0 & 12.0 &  -8.5 &  8.5 \\
\hline
\end{tabular}
\end{center}
\caption{\it Axial and correlated limits for the $Z\g\g$ anomalous couplings
at HERA with different integrated luminosities and $95\%$ CL. \label{table}}
\end{table}

The origin of so poor results is the fact that, unlike diboson production
at hadron or $e^+e^-$ colliders, the anomalous diagrams of $ep\to e\g X$
have a $Z$ propagator decreasing their effect. 
The process $ep\to eZX$ avoids this problem thanks to the absence
of these propagators: the Standard Model cross section is similar
to the anomalous one but, as a drawback, they are of the order of
femtobarns.

\section{Summary and conclusions}

The radiative neutral current process $ep\to e\g X$ 
at HERA has been studied. Realistic cuts have been applied in order to
observe a clean signal consisting of detectable and well separated 
electron, photon and jet. 

The possibility of testing the trilinear neutral gauge boson couplings
in this process has also been explored. The $Z\g Z$ couplings are
very suppressed by two $Z$ propagators. Only the $Z\g \g$ couplings
have been considered. A Monte Carlo program has been developed to
account for such anomalous vertex and further cuts have been implemented 
to improve the sensitivity to this source of new physics.
Our estimates are based on total cross sections since the expected number
of events is so small that a distribution analysis is not possible.
The distributions just helped us to find the optimum cuts. Unfortunately,
competitive bounds on these anomalous couplings cannot be achieved at
HERA, even with the future luminosity upgrades.\footnote{We would like to
apologize for the optimistic but incorrect results that were presented
at the workshop due to a regrettable and unlucky mistake in our programs.}
As a counterpart, a different kinematical region is explored, in which
the form factors can be neglected.

\section*{Acknowledgements}

One of us (J.I.) would like to thank the Workshop organizers for financial
support and very especially the electroweak working group conveners 
and the Group from Madrid at ZEUS for hospitality and useful conversations. 
This work has been partially supported by the CICYT and the European Commission
under contract CHRX-CT-92-0004.


\end{document}